# Stacking-dependent ferroicity of reversed bilayer: altermagnetism or ferroelectricity

Wencong Sun[1], Haoshen Ye[1], Li Liang[1], Ning Ding[1], Shuai Dong,[1, *] and Shan-shan Wang[1, †]

[1]*Key Laboratory of Quantum Materials and Devices of Ministry of Education, School of Physics, Southeast University, Nanjing 211189, P. R. China.*

**Abstract**

Altermagnetism, as a new branch of magnetism independent of traditional ferromagnetism and antiferromagnetism, has attracted extensive attention recently. At present, researchers have proved several kinds of three-dimensional altermagnets, but research on two-dimensional (2D) altermagnets remains elusive. Here, we propose a method for designing altermagnetism in 2D lattices: bilayer reversed stacking. This method could enable altermagnetism-type spin splitting to occur intrinsically and the spin-splitting can be controlled by crystal chirality. We also demonstrate it through a real material of bilayer $PtBr_3$ with AB' stacking order. Additionally, the combination of stacking order and slidetronics offers new opportunities for electrical writing and magnetic reading of electronic devices. In the case of AC' stacking, interlayer sliding results in reversible spontaneous polarization. This unique combination of antiferromagnetism and sliding ferroelectricity leads to polarization-controlled spin-splitting, thus enabling magnetoelectric coupling, which can be detected by magneto-optical Kerr effect even without net magnetization. Our research highlights that reversed stacking provides a platform to explore rich physical properties of magnetism, ferroelectricity, and spin-splitting.

## Introduction

Engineering and manipulating spins of electrons in space, momentum, and energy have become the frontier research in spintronics. Conventional spintronic devices utilize ferromagnets as spin generators and manipulators [1,2]. Recently, the study of antiferromagnets-based devices brought new developments in spintronics [3,4]. Compared with ferromagnets, antiferromagnetic materials have high information storage density and unique terahertz (Thz) spin dynamics, which enable magnetic moment reversal on picosecond time scales [5]. However, the application of antiferromagnets remains challenging. Antiferromagnets are much less sensitive to external magnetic field and it is hard to manipulate the spin orders. Thus, an ideal next-generation spintronics needs to have the properties of easy manipulation as well as the ability to store information with high stability, high density, and ultrafast spin dynamics. Such systems can be realized in spin-splitting antiferromagnets. The spin-splitting can arise from Zeeman effect [6], Rashba [7,8] and Dresselhaus [9] spin-orbit coupling (SOC), and external field effects.

Very recently, altermagnetism, a new branch of magnetism independent of traditional ferromagnetism (FM) and antiferromagnetism (AFM), has been proposed [10–17]. The altermagnetism has both spin polarization observed in ferromagnets and antiparallel spin arrangement with vanishing net magnetization found in collinear antiferromagnets. The typical feature of altermagnetism is the band structure displays alternating momentum-dependent spin-splitting even without SOC.

The advent of 2D materials has revolutionized the field of material science, offering promising applications in miniaturized electronic devices [18–21]. The weak van der Waals interaction between layers of 2D materials makes it possible to regulate order parameters by stacking. The 2D stacked bilayers provide a platform for exploring novel quantum states such as ferroelectricity (FE), magnetism, and topology [22–27]. These phases are markedly sensitive to the relative stacking orders of the layers, providing a powerful knob for tailoring material properties. For example, nonpolar monolayers can be stacked into polar bilayer structures, enabling ferroelectric polarization switching through subtle lateral shifts between layers [28–30]. With the rise of sliding physics, the ferromagnetic monolayer can be stacked into bilayer antiferromagnets. Based on the paradigm of mediating sliding physics, the crystalline symmetry and magnetic orders change and thus give rise to spin-splitting and anomalous Hall effect [31–34] in those antiferromagnetic bilayer systems. Actually, the spin-splitting in antiferromagnets has not been thoroughly explored in bilayer lattices.

In this Letter, we investigate the altermagnetic phase, magnetic spin-splitting and ferroelectric properties of $PtBr_3$ bilayers by symmetry analysis and first-principles calculations. The $PtBr_3$ bulk was synthesized for more than a century [35–37], and recently its monolayer counterpart has been researched [38–40]. Here, we research the bilayer $PtBr_3$ with reversed stacking orders, i.e. the upper single layer is stacked on the lower layer after $M_z$ mirror symmetry operation. Different from ordinary stacked double-layers, reversed stacking can break the *space inversion* symmetry and may produce novel phenomena. We investigate the ferroelectric, magnetic, and anomalous Hall effect of $PtBr_3$ bilayer under three reversed stacking orders, AA', AB' and AC'. For the AB' stacking, the altermagnetism is constructed due to the symmetry breaking, accompanied by chirality-reversible spin-splitting and crystal Hall effect [41–43]. For the AC' stacking, ferroelectric polarization appears in the antiferromagnetic bilayer, leading to polarization-controlled spin-splitting, thus enabling magnetoelectric coupling, which can be detected by magneto-optic Kerr effect (MOKE) even with zero magnetization. The tunable spin-splitting from the stacking orders may provide a platform for designing spintronic devices.

**Methods**

Based on density functional theory (DFT), the first principles calculations were conducted using Vienna *ab-initio* Simulation Package (VASP) with the projector augmented-wave (PAW) method [44]. The kinetic energy cutoff of the plane wave basis was set to 500 eV. The exchange-correlation energy was approximated using the generalized gradient approximation (GGA) as parametrized by Perdew, Burke, and Ernzerhof [45]. The energy convergence criterion was set to $10^{-5}$ eV and the forces were converged to less than 0.01 eV/Å on each atom. The Brillouin zone integration was sampled using a Γ-centered $9 \times 9 \times 1$ $k$-point mesh. After testing the DFT+$U$ correction, it was found that Hubbard $U_{eff}$ = 1 eV was appropriate to deal with electron correlation of Pt's $5d$ orbitals [46]. A vacuum layer of more than 15 Å was used to avoid periodic potential along the out-of-plane direction. The van der Waals correction was properly described by DFT-D3 method [47]. The maximally localized Wannier functions were constructed using the WANNIER90 package [48] by projecting onto *p* orbital of Br atom and *d* orbital of Pt atom to calculate the MOKE. When calculating Kerr angle and ellipticity, we chose $SiO_2$ as a 2D materials substrate with refractive index *n*=1.546. Berry curvature and anomalous Hall conductance were calculated using the WANNIER90 [48] and Wanniertools packages [49]. The energy barrier of polarization switching was calculated by climbing image nudged elastic band (CI-NEB) method [50].

**Results and Discussion**

The crystal structure of $PtBr_3$ monolayer is depicted in Fig. 1(a). The Pt cation and the six surrounding Br anions form an octahedron, whose (111) direction points to the out-of-plane direction (i.e. the *c*-axis). Generally, the octahedral crystal field splits the *d* orbitals into the low-lying $t_{2g}$ triplets ($d_{xz}/d_{yz}/d_{xy}$) and higher-energy $e_g$ doublets ($d_{z2}/d_{x2-y2}$). However, due to the trigonal distortion of the regular octahedron by compression along the *c*-axis, the $t_{2g}$ triplets split into the low-energy $a_{1g}$ singlet and high-energy degenerate $e_g'\pm$, and $e_g$ orbitals split into two degenerate $e_g\pm$ [51]. The $Pt^{3+}$ with $5d^7$ electron configuration has two possibilities of orbital occupation: high and low spin states. The orbital occupation in the high spin state is $a_{1g}^2\ e_g'^2_-\ e_g'^1_+\ e_{g-}^1\ e_{g+}^1$ with 3 $\mu_B$ magnetic moment, where the half-filled occupation of orbital results in the metallic state. While, the orbital occupation of the low spin state is $a_{1g}^2\ e_g'^2_-\ e_g'^2_+\ e_{g-}^1\ e_{g+}^0$ with 1$\mu_B$ magnetic moment, corresponding to the insulation state. Our calculation results show that the low spin state with 1 $\mu_B$/Pt ion is the ground state with a lower energy. The fully optimized lattice constant of monolayer is 6.732 Å, consistent with the previous calculation [38].

Here, we consider three stacking orders with the top layer reversed by a $M_z$ mirror symmetry operation with respect to the bottom layer, which breaks the space inversion symmetry. Specifically, the AA' stacking order is constructed by placing one layer on top of another layer after, as shown in Fig. 1(b). The stacking orders of AB' and AC' bilayers are constructed based on the AA' stack by sliding the top layer horizontally relative to the bottom layer in fractional coordinates (2/3, 1/3, 0) and (1/3, 0, 0), respectively, as illustrated in Figs.

1(c) and 1(d). The energy changes of the three reversed stacking modes during the interlayer sliding process are shown in Fig. S1.

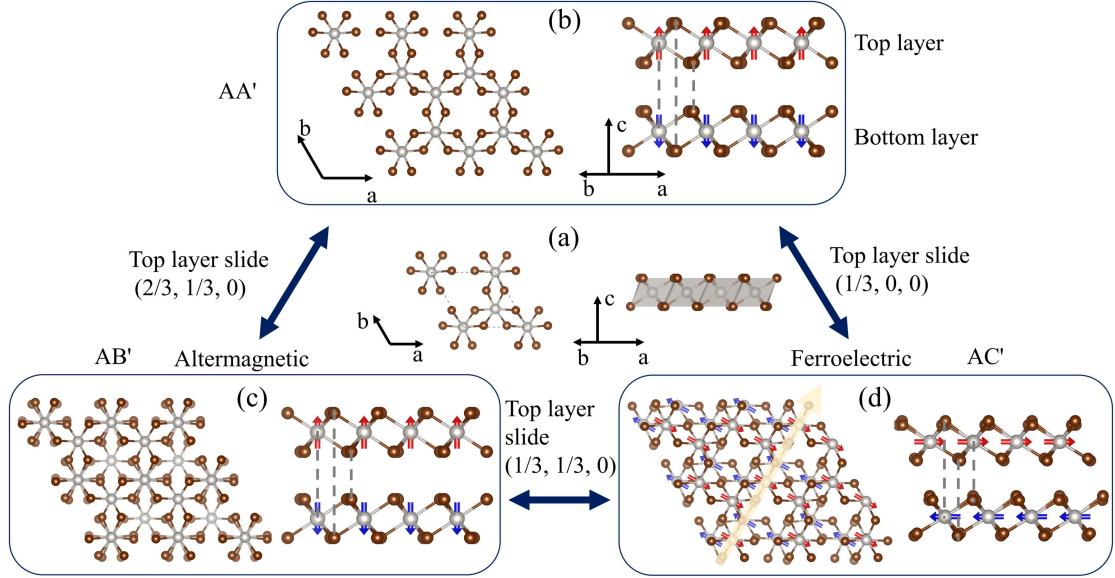

FIG. 1. (a) The top and side view of PtBr$_3$ monolayer. (b), (c), and (d) are AA', AB', and AC' stacking orders, respectively. The gray dashed lines provide a visual guide for the side view. The red and blue arrows indicate the orientation of spin up and down, respectively. In (d), the yellow arrow shows the direction of in-plane polarization

TABLE I. The information for stacking bilayers, including energy difference of different stacking orders and magnetic orders with AC' magnetic ground state as reference (in units of meV/f.u.), the orientation of the magnetic easy-axis/plane and band gaps without/with consideration of SOC (in units of meV).

|  | FM | A-AFM | Néel | Stripy | Zigzag | Easy axis/plane | Band gap |
|---|---|---|---|---|---|---|---|
| AA' | 122.4 | 106.1 | 122.9 | 125.8 | 125.7 | $z$ | 0/11 |
| AB' | 85.3 | 80.1 | 92.3 | 94.6 | 95.8 | $z$ | 378/144 |
| AC' | 9.5 | 0.0 | 24.7 | 6.0 | 21.7 | $x$-$y$ | 440/414 |

The magnetic properties of these stacking orders are investigated by first-principles calculations. Several magnetic orders have been considered, including ferromagnetism (FM), A-AFM, Néel-AFM, Stripy-AFM and Zigzag-AFM, where AFM magnetic configurations are shown in Fig. S2. The results are summarized in Table 1. The magnetic ground states of all these stacking favor the A-AFM magnetic order. The magnetocrystalline anisotropy energies have also been calculated by rotating the spin magnetic moment along various directions. The results are depicted in Fig. S3. It is worth noting that the easy magnetization directions for AA' and AB' bilayers are oriented along the $z$-axis, while the easy magnetization axis of AC' stack lies in the $x$-$y$ plane.

Symmetry plays an important role in AFM systems. Certain effects are typically forbidden by time inversion, space inversion, space translation, spinor reversal symmetry, as well as their combined symmetry operation [14]. Hence, AFM materials usually behave like non-magnetic materials in many ways, which are hard to manipulate by electrical or optical techniques [14]. Considering the nonrelativistic phenomenology of the bilayer stacking phases, the spin space and real space are decoupled. Therefore, the stacking phases can be described using spin layer group. Previous research have indicated that the spin group formalism can accurately distinguish ferromagnetic, antiferromagnetic and altermagnetic phases. [11,52]. Spin groups can be represented as the direct product $r_s \otimes R_s$, where $r_s$ refers to the spin-only group, consists only of spin space transformations, and $R_s$ refers to nontrivial spin groups that contain pairs of transformations $[R_i \| R_j]$ The transformation on the left of $[R_i \| R_j]$ acts solely on the spin space, while the transformation on the right of the double bar acts exclusively on the real space [52]. The symmetric operations in real space of the three stacking orders are shown in Table 2. For AA' stacking, symmetry operations containing three-fold rotation $C_{3z}$ and double-fold rotation $C_{21}$, $C_{22}$, $C_{23}$ link the atoms within the same spin. While, symmetry operations containing $6z$ and $m_z$, $m_x$, $m_y$, $m_{xy}$ link the atoms with opposite spin. From the perspective of spin group, since AA' stacking has $m_z$ symmetry, the nontrivial spin layer group contains the $[C_2 \| m_z]$ symmetry transformation. When we apply it to spin and momentum dependent bands, we can obtain $[C_2 \| m_z]\varepsilon(s,k) = \varepsilon(-s,k)$. Since the $[C_2 \| m_z]$ is the symmetry operation of the system, we have $[C_2 \| m_z]\varepsilon(s,k) = \varepsilon(-s,k) = \varepsilon(s,k)$. Hence, the spin degeneracy will appear in AA' stacking.

For AB' stacking, the symmetry operations are {$E$, $C_{3z}^+$, $C_{3z}^-$, $C_{2x}$, $C_{2y}$, $C_{2xy}$}. The $C_{3z}$ symmetry connects atoms within the same spin, while the $C_{2x}$ and $C_{2y}$ and $C_{2xy}$ connect atoms with opposite spins. The spin layer group is given by $R_3 = [E \| H] + [C_2 \| G - H]$. The spin degeneracy of the band structures can be protected by the $[C_2 \| G - H]$ symmetries. For example, the $[C_2 \| C_{2x}]$ protects the spin degeneracy along the G-M (where $k_y$=0). The spin degeneracy of band structure along high symmetry lines G-K and K-M are protected by the $[C_2 \| C_{2y}]$ and $[C_2 \| C_{2xy}]$, respectively. The layer group of AB' stacking are 68. Consequently, its spin layer group is identified as $^1\bar{3}^2m$ with hexagonal characteristic of spin-momentum locking [52]. Our DFT analysis are consistent with the symmetry analysis using spin layer group, as shown in Fig.2 (d,e).

For AC' stacking, due to the existence of electric dipole and potential difference, the antiferromagnetic configuration produces an uncompensated non-zero total magnetic moment. The symmetry operation only contains a vertical mirror symmetry $m_y$. Hence, the spin layer group of AC' stacking belongs to the spin group that describes the nonzero magnetization phases.

The electronic band structure of AB' calculated by DFT is consistent with the above symmetry analysis. As illustrated in Fig. 2(b), along the non-highly symmetric the K-K2 path, there is a distinct separation between the spin-up and spin-down bands, while spin degeneracy remains along the highly symmetric paths K-Γ, Γ-M and M-K. The spin-splitting projections [Figs. 2(d-e)] unambiguously illustrate spin-splitting distribution characterized by alternating momentum-dependent signs. As depicted in Figs. 2(d) and 2(e), spin-splitting energies at the highest valence band (HVB) and lowest conduction band (LCB) reach up to 5.5 meV and 18 meV, respectively. The significant spin-splitting energies are supposed to be directly observed in experiments through spin-and-angle-resolved photoemission spectroscopy (SARPES).

It is noted that under the symmetry constraints, we obtain two energetically equivalent stacking patterns, i.e., AB'1 and AB'2 as demonstrated in Fig. 2(f). For the AB'2 pattern, it can be obtained from AB'1 under a lattice inversion symmetry. Therefore, the two AB' patterns have opposite chirality and the spin density, which gives rise to a reversible spin-splitting, as shown in Figs. 2(b) and 2(c).

TABLE II. The symmetry analysis for different stacking orders, including magnetic point

group (MPG) and symmetry operations.

|  | Symmetry operations in real space |
|---|---|
| AA' | $3_z, 3_z^{-1}, 2_1, 2_2, 2_3, -6z, -6z^{-1}, m_z, m_x, m_{zy}, m_y$ |
| AB' | $3_z, 3_z^{-1}, 2_x, 2_{xy}, 2_y$ |
| AC' | $m_y$ |

The combination of time-reversal and spatial inversion symmetry *PT* excludes the Hall conductivity in collinear antiferromagnets. Unlike the *PT* symmetries, the two-fold rotation symmetry does not necessarily prohibit the Hall currents. Thus, the anomalous Hall effect may occur in both AB'1 and AB'2 stacking orders, even though the net magnetization vanishes. DFT calculation results show that there are AHE conductance signals emergent as shown in Fig. 2(f). Thus, the altermagnetic properties can be experimentally detected using anomalous Hall effect (AHE) measurements [41,42]. What's more, the inversion of crystal chirality can result in the opposite sign of the Hall conductance [Fig. 2(f)], which is known as the crystal Hall effect [41–43]. The proposed design principles of reversible altermagnetism-type spin-splitting and crystal Hall effect can be generalized to other $MX_3$ materials, for example, AB' stacking order of MnBr$_3$ bilayer has been predicted to exhibit altermagnetic as well, as shown in Fig. S4 in the supplementary [53].

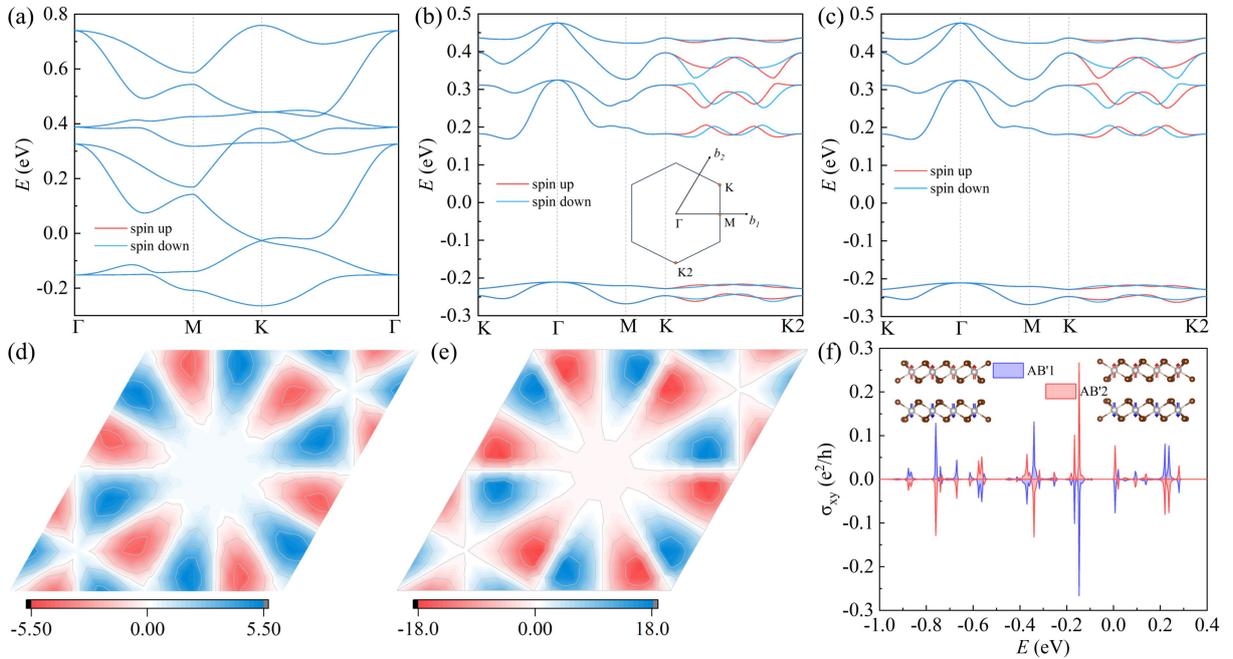

FIG. 2. (a) The band structure of AA' stacking order without SOC. (b-f) The electronic properties of AB' stacking order. (b) and (c) are band structures without consideration of SOC

for AB'1 and AB'2, where the inset of (b) shows 2D Brillouin zone and high symmetry points. (d) and (e) depict spin-splitting energy projections (in units of meV) of HVB and LCB respectively for AB'1, where the pink (blue) color block indicates $E_{\text{spin-up}} > E_{\text{spin-down}}$ ($E_{\text{spin-up}} < E_{\text{spin-down}}$). (f) Crystal Hall effect. The illustrations show AB'1 and AB'2 structures.

The interlayer sliding of 2D materials can effectively modulate the symmetry and electronic environment, potentially leading to spontaneous polarization and enabling new functionalities. In the AC' bilayer, the interlayer sliding breaks spatial inversion symmetry, generating a switchable electric polarization of 2.4 pC/m along the out-of-plane (OOP) direction as shown in Fig. 3(a), which is comparable to that of the *h*-BN bilayer (2.08 pC/m) [28] and much larger than that of the WTe$_2$ (0.38 pC/m) [30] and MoS$_2$ bilayer (0.97 pC/m) [28]. In addition to OOP polarization, AC' stack also exhibits in-plane (IP) polarization with the value of 9.7 pC/m, where the polarization direction is shown in Fig. 1(d). In order to determine the most favorable path of vertical FE flipping, the energy contour plot with lateral translation is depicted in Fig. S6. where the dashed arrowed path represents the most likely trajectory. The energy barrier calculations on this path, as shown in Fig. 3(a), indicate that the FE flip only needs to overcome 18 meV/f.u., which is much lower than that of the traditional ferroelectric barrier, but higher than the BN bilayer (0.15 meV/f.u.) [28] and WTe$_2$ bilayer (4.5 meV/f.u.) [30]. The structures of the initial, intermediate and final states are shown in the inset of Fig. 3(a). Notably, the intermediate state exhibits no net out-of-plane polarization, while in-plane polarization persists.

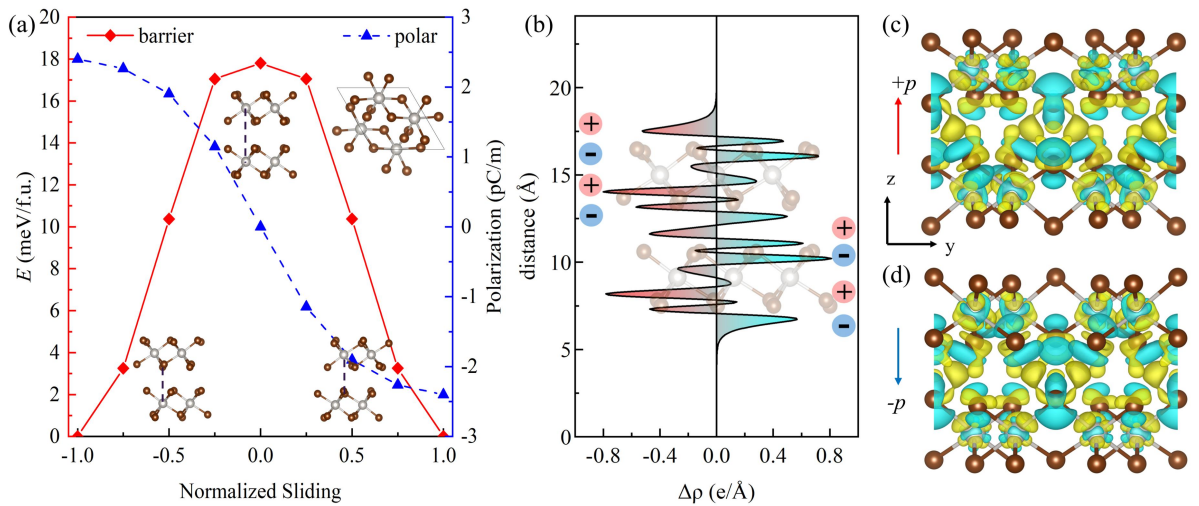

FIG. 3. (a) The variation of energy and OOP polarization as a function of normalized sliding displacement. The illustrations represent the initial, intermediate, and final state structures during interlay movement. (b) The planar-integrated differential charge density as the function of $z$-axis distance, with $-P$ phase as the reference phase. The positive values of charge density ($\Delta\rho$) represent electrons and negative values represent holes. The red and blue spheres indicate positively and negatively charged areas, respectively. (c) and (d) are differential charge density plots of $+P$ and $-P$ phases, respectively, where the yellow and lake green areas represent the gained and lost electrons with the isovalue of $1.82\times10^{-4}$ e/Å$^3$, respectively.

To elucidate the origin of OOP ferroelectric polarization, we conducted the planar-integrated differential charge density calculation along the $z$-axis, which defined as the difference between the charge densities of the $+P$ and $-P$ phases. As shown in Fig. 3(b), the positive values represent regions where there is an excess of electrons, while the negative values indicate regions where there is a deficit of electrons, commonly referred to as holes. For each PtBr$_3$ monolayer, due to the uneven charge distribution, dipoles with $+z$ direction are formed in both the upper and lower sublayers. This phenomenon occurs because the sliding motion breaks the inversion symmetry between the upper and lower Pt-Br bonds, altering their chemical environments, resulting in one side of the material becoming more positively charged while the other side accumulates more negative charges. Due to the top and bottom monolayers having dipoles along the $+z$ direction, bilayer forms a net positive polarization ($+P$), which is consistent with the DFT calculation. In addition, it is obvious from the charge-differential density diagrams that the local positive and negative charge distributions between the $+P$ phase and the negative polarization ($-P$) phase are opposite, as shown in Figs. 3(c) and 3(d).

A recent study has shown that A-AFM can produce global spin-splitting in momentum space under the applied external electric field [54]. Ferroelectric polarization results in electrostatic potential difference between layers, forming a built-in electric field, which disrupts the spin degeneracy of the antiferromagnet and triggers spontaneous spin-splitting in the whole momentum space. The spin up and spin down channels are flipped by sliding FE, producing polarization-controlled spin-splitting. The band structures without SOC of $+P$ and -

$P$ phase for AC' stack are shown in Figs. 4(a) and 4(b), respectively. It can be seen that the spin-splitting is reversed upon switching the polarization, which illustrates the magnetoelectrical coupling.

The projected band maps including SOC display that the top and bottom layer bands near Fermi energy are separate. For +$P$ phase, the conduction band minimum (CBM) [valence band maximum (VBM)] band is mainly contributed by the top (bottom) layer, as shown in Fig. 4(c). On the contrary, in the -$P$ phase, the CBM (VBM) band is primarily from the bottom (top) layer, as shown in Fig. 4(d). Simultaneously, Berry curvatures reverse with switching polarization by interlayer sliding, as shown in Fig. S7(a-b) of the supplementary [50], presenting layer-locked characteristics. The layer-locked Berry curvatures induce layer-polarized anomalous Hall effect (LPAHE)[31-34], in which electrons from the top and bottom layers spontaneously deflect in opposite directions. The anomalous Hall conductance calculations further prove the LPAHE phenomenon of AC', as shown in Fig. S7(c) of the supplementary [50].

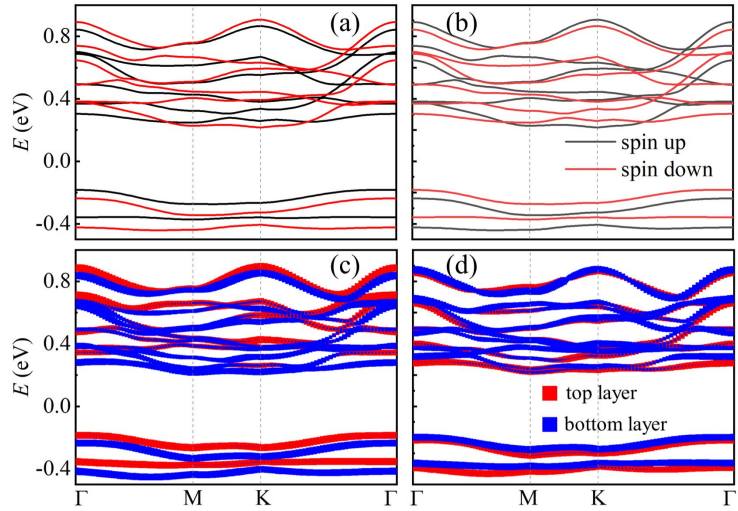

FIG. 4. The band structures of +$P$ (a, c) and -$P$ (b, d) phases for AC' stack. (a-b) are band structures without SOC. (c-d) are layer-resolved band structures with consideration of SOC.

Next, we confirm the regulation of polarization on magnetism through MOKE simulation in the AFM. The MOKE refers to the change in the polarization state of light reflected from a magnetized surface, enabling the direct detection of magnetic properties. In general, the MOKE signal can be detected in antiferromagnetic materials with broken *PT* symmetries [55–57]. Due to the reversed stacking configuration with inherent broken *PT* symmetry, the AC' order can activate MOKE signal without applying an external field or constructing complex heterojunction. It is important to note that the MOKE, characterized by changes in the Kerr angle and ellipticity, exhibits a reversal between +*P* phase and -*P* phase, as illustrated in Figs. 5(a) and 5(b). The calculation formulas and details are in the supplementary. To sum up, our findings not only indicate the feasibility of rendering "electrical writing and magnetic reading" but also provide a way to use MOKE optics to extract the information stored in polar antiferromagnetic states.

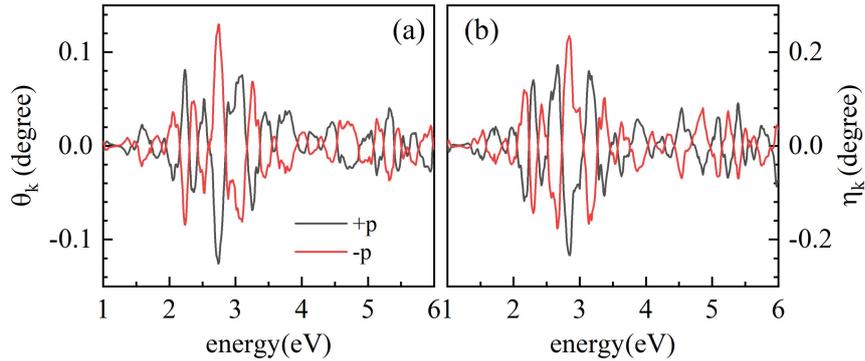

FIG. 5. The Magneto-optical response of the +*P* and -*P* phases for AC' order, characterized by Kerr angle (a) and ellipticity (b).

**Conclusion**

In conclusion, we explored the unique properties of bilayer PtBr$_3$ with reversed stacking orders, specifically focusing on ferroelectric, magnetic, and spin-splitting phenomena. Our findings reveal that different stacking orders significantly influence the magnetic and ferroelectric characteristics. The AB' stacking shows considerable chirality-reversible spin-splitting with alternating momentum-dependent signs in the band structure, opening pathways for designing altermagnetism in similar M$X_3$ structures. At the same time, a new type of crystal Hall effect has been found in AB' stacking. AC' stacking bilayer exhibits both in-plane and switchable out-of-plane spontaneous FE polarizations. The coexistence of polarization and antiferromagnetism induces polarization-controllable spin-splitting in the whole momentum space, LPAHE phenomena, and multiferroic properties detectable by MOKE. Our research highlights that reversed stacking provides a platform to explore the rich physical properties of magnetism, altermagnetism, ferroelectricity, and multiferroics.

**Author information**

Corresponding Authors

*Email: sdong@seu.edu.cn

†Email: wangss@seu.edu.cn

**Acknowledgments**

This work is supported by the Natural Science Foundation of China (Grant Nos. 12104089, 12325401, 12274069), Postgraduate Research & Practice Innovation Program of Jiangsu Province (Grant No. KYCX23_0222). Most calculations were done on the Big Data Computing Center of Southeast University.

*Note added in proof.* Very recently, Pan *et al.* [58] and Zeng *et al.* [59] independently studied bilayer stacking altermagnetism. Despite the shared focus on bilayer stacking altermagnetism, our work differs from these studies in content: we focused on altermagnetism induced by the reversed stacking operation.

**References**

[1] T. Dietl and H. Ohno, Dilute ferromagnetic semiconductors: Physics and spintronic structures, Rev. Mod. Phys. **86**, 187 (2014).

[2] A. Hirohata, K. Yamada, Y. Nakatani, I.-L. Prejbeanu, B. Diény, P. Pirro, and B. Hillebrands, Review on spintronics: Principles and device applications, J. Magn. Magn. Mater. **509**, 166711 (2020).

[3]  T. Jungwirth, X. Marti, P. Wadley, and J. Wunderlich, Antiferromagnetic spintronics, Nat. Nanotechnol. **11**, 231 (2016).

[4]  V. Baltz, A. Manchon, M. Tsoi, T. Moriyama, T. Ono, and Y. Tserkovnyak, Antiferromagnetic spintronics, Rev. Mod. Phys. **90**, 015005 (2018).

[5]  K. Olejník et al., Terahertz electrical writing speed in an antiferromagnetic memory, Sci. Adv. **4**, eaar3566 (2018).

[6]  H. Yuan et al., Zeeman-type spin splitting controlled by an electric field, Nat. Phys. **9**, 563 (2013).

[7]  Y. A. Bychkov and E. I. Rashba, Properties of a 2D electron gas with lifted spectral degeneracy, JETP Lett. **39**, 78 (1984).

[8]  F. G. Pikus and G. E. Pikus, Conduction-band spin splitting and negative magnetoresistance in A3B5 heterostructures, Phys. Rev. B **51**, 16928 (1995).

[9]  G. Dresselhaus, Spin-orbit coupling effects in zinc blende structures, Phys. Rev. **100**, 580 (1955).

[10] R. He, D. Wang, N. Luo, J. Zeng, K. Q. Chen, and L. M. Tang, Nonrelativistic spin-momentum coupling in antiferromagnetic twisted bilayers, Phys. Rev. Lett. **130**, 046401 (2023).

[11] L. Šmejkal, J. Sinova, and T. Jungwirth, Beyond conventional ferromagnetism and antiferromagnetism: A Phase with nonrelativistic spin and crystal rotation symmetry, Phys. Rev. X **12**, 031042 (2022).

[12] S. Lee et al., Broken Kramers degeneracy in altermagnetic MnTe, Phys. Rev. Lett. **132**, 036702 (2024).

[13] S. Bhowal and N. A. Spaldin, Ferroically ordered magnetic octupoles in d-wave altermagnets, Phys. Rev. X **14**, 011019 (2024).

[14] L. Šmejkal, J. Sinova, and T. Jungwirth, Emerging research landscape of altermagnetism, Phys. Rev. X **12**, 040501 (2022).

[15] X. Zhou, W. Feng, R. W. Zhang, L. Šmejkal, J. Sinova, Y. Mokrousov, and Y. Yao, Crystal thermal transport in altermagnetic RuO2, Phys. Rev. Lett. **132**, 056701 (2024).

[16] J. Krempaský et al., Altermagnetic lifting of Kramers spin degeneracy, Nature **626**, 517 (2024).

[17] I. I. Mazin, K. Koepernik, M. D. Johannes, R. González-Hernández, and L. Šmejkal, Prediction of unconventional magnetism in doped FeSb2, Proc. Natl. Acad. Sci. U. S. A. **118**, e2108924118 (2021).

[18] C. Gong and X. Zhang, Two-dimensional magnetic crystals and emergent heterostructure devices, Science **363**, eaav4450 (2019).

[19] K. S. Burch, D. Mandrus, and J.-G. Park, Magnetism in two-dimensional van der Waals materials, Nature **563**, 47 (2018).

[20] D. Zhang, P. Schoenherr, P. Sharma, and J. Seidel, Ferroelectric order in van der Waals layered materials, Nat. Rev. Mater. **8**, 25 (2023).

[21] M. An and S. Dong, Ferroic orders in two-dimensional transition/rare-earth metal halides, APL Mater. **8**, 110704 (2020).

[22] Y. Jiao, X. T. Zeng, C. Chen, Z. Gao, K. Song, X. L. Sheng, and S. A. Yang, Monolayer and bilayer PtCl3: energetics, magnetism, and band topology, Phys. Rev. B **107**, 075436 (2023).

[23] X. Kong, H. Yoon, M. J. Han, and L. Liang, Switching interlayer magnetic order in bilayer CrI3 by stacking reversal, Nanoscale **13**, 16172 (2021).

[24] W. Xun, C. Wu, H. Sun, W. Zhang, Y. Z. Wu, and P. Li, Coexisting magnetism, ferroelectric, and ferrovalley multiferroic in stacking-dependent two-dimensional materials, Nano Lett. **24**, 3541 (2024).

[25] N. Sivadas, S. Okamoto, X. Xu, C. J. Fennie, and D. Xiao, Stacking-dependent magnetism in bilayer CrI3, Nano Lett. **18**, 7658 (2018).

[26] X. Li, X. Xu, H. Zhou, H. Jia, E. Wang, H. Fu, J. T. Sun, and S. Meng, Tunable

topological states in stacked chern insulator bilayers, Nano Lett. **23**, 2839 (2023).

[27] T. Cao, D. F. Shao, K. Huang, G. Gurung, and E. Y. Tsymbal, Switchable anomalous hall effects in polar-stacked 2D antiferromagnet MnBi2Te4, Nano Lett. **23**, 3781 (2023).

[28] L. Li and M. Wu, Binary compound bilayer and multilayer with vertical polarizations: two-dimensional ferroelectrics, multiferroics, and nanogenerators, ACS Nano **11**, 6382 (2017).

[29] N. Ding, J. Chen, C. Gui, H. You, X. Yao, and S. Dong, Phase competition and negative piezoelectricity in interlayer-sliding ferroelectric ZrI2, Phys. Rev. Mater. **5**, 084405 (2021).

[30] Q. Yang, M. Wu, and J. Li, Origin of two-dimensional vertical ferroelectricity in WTe2 bilayer and multilayer, J. Phys. Chem. Lett. **9**, 7160 (2018).

[31] A. Gao et al., Layer Hall effect in a 2D topological axion antiferromagnet, Nature **595**, 521 (2021).

[32] R. Peng, T. Zhang, Z. He, Q. Wu, Y. Dai, B. Huang, and Y. Ma, Intrinsic layer-polarized anomalous Hall effect in bilayer MnBi2Te4, Phys. Rev. B **107**, 085411 (2023).

[33] T. Zhang, X. Xu, B. Huang, Y. Dai, L. Kou, and Y. Ma, Layer-polarized anomalous Hall effects in valleytronic van der Waals bilayers, Mater. Horizons **10**, 483 (2022).

[34] T. Zhang, X. Xu, J. Guo, Y. Dai, and Y. Ma, Layer-polarized anomalous Hall effects from inversion-symmetric single-layer lattices, Nano Lett. **24**, 1009 (2024).

[35] H. G. Von Schnering, J. H. Chang, M. Freiberg, K. Peters, E. M. Peters, A. Ormeci, L. Schröder, G. Thiele, and C. Röhr, Structure and bonding of the mixed-valent platinum trihalides, PtCl3 and PtBr3, Z. Anorg. Allg. Chem. **630**, 109 (2004).

[36] P. W. D.-C. Priv.-Doz. Dr. G. Thiele, Platin(III)-bromid – ein neuer Strukturtyp von AB3-Verbindungen, Angew. Chem. 706 (1969).

[37] D. binären B. und J. L. Wöhler and F. Müller and D. Platins, Die Binären Bromide und Jodide des Platins, Z. Anorg. Allg. Chem. **149**, 377 (1925).

[38] J. Y. You, Z. Zhang, B. Gu, and G. Su, Two-dimensional room-temperature ferromagnetic semiconductors with quantum anomalous hall effect, Phys. Rev. Appl. **12**, 024063 (2019).

[39] X. Xu, Z. Sun, X. Wang, Z. Gao, L. Guan, S. Zhang, P. Chang, and J. Tao, Tunable magnetic coupling and high Curie temperature of two–dimensional PtBr3 via van der waals heterostructures, Appl. Surf. Sci. **572**, 151478 (2022).

[40] J. Y. You, X. J. Dong, B. Gu, and G. Su, Possible Room-Temperature Ferromagnetic Semiconductors, Chinese Phys. Lett. **40**, 067502 (2023).

[41] L. Šmejkal, L. Šmejkal, L. Šmejkal, R. González-Hernández, R. González-Hernández, T. Jungwirth, T. Jungwirth, J. Sinova, and J. Sinova, Crystal time-reversal symmetry breaking and spontaneous Hall effect in collinear antiferromagnets, Sci. Adv. **6**, eaaz8809 (2020).

[42] Z. Feng et al., An anomalous Hall effect in altermagnetic ruthenium dioxide, Nat. Electron. **5**, 735 (2022).

[43] L. Šmejkal, A. H. MacDonald, J. Sinova, S. Nakatsuji, and T. Jungwirth, Anomalous Hall antiferromagnets, Nat. Rev. Mater. **7**, 482 (2022).

[44] G. Kresse and D. Joubert, From ultrasoft pseudopotentials to the projector augmented-wave method, Phys. Rev. B **59**, 1758 (1999).

[45] J. P. Perdew, K. Burke, and M. Ernzerhof, Generalized gradient approximation made simple, Phys. Rev. Lett. **77**, 3865 (1996).

[46] S. L. Dudarev, G. A. Botton, S. Y. Savrasov, C. J. Humphreys, and A. P. Sutton, Electron-energy-loss spectra and the structural stability of nickel oxide: An LSDA+U study, Phys. Rev. B **57**, 1505 (1998).

[47] S. Grimme, J. Antony, S. Ehrlich, and H. Krieg, A consistent and accurate ab initio

parametrization of density functional dispersion correction (DFT-D) for the 94 elements H-Pu, J. Chem. Phys. **132**, 154104 (2010).

[48] A. A. Mostofi, J. R. Yates, Y.-S. Lee, I. Souza, D. Vanderbilt, and N. Marzari, Wannier90: A tool for obtaining maximally-localised Wannier functions, Comput. Phys. Commun. **178**, 685 (2008).

[49] Q. Wu, S. Zhang, H.-F. Song, M. Troyer, and A. A. Soluyanov, WannierTools: An open-source software package for novel topological materials, Comput. Phys. Commun. **224**, 405 (2018).

[50] G. Henkelman, B. P. Uberuaga, and H. Jónsson, A climbing image nudged elastic band method for finding saddle points and minimum energy paths, J. Chem. Phys. **113**, 9901 (2000).

[51] S. Landron and M. B. Lepetit, Importance of t2g-eg hybridization in transition metal oxides, Phys. Rev. B **77**, 125106 (2008).

[52] S. Zeng and Y. J. Zhao, Description of two-dimensional altermagnetism: Categorization using spin group theory, Phys. Rev. B **110**, 54406 (2024).

[53] See Supplemental Material at http://link.aps.org/supplemental/10.1103/PhysRevB.xxx.xxxxxx for the DFT-calculated lattice constants of bulk $PtBr_3$, in comparison with experimental result; the spin-splitting in electric band structure of $MnBr_3$; contour plot of Berry curvature of AB' order at valence band maximum; the Berry curvatures and anomalous Hall conductance for AC' stacking; and calculation formula and details of MOKE.

[54] H. Lv, Y. Niu, X. Wu, and J. Yang, Electric-field tunable magnetism in van der Waals bilayers with A-type antiferromagnetic order: unipolar versus bipolar magnetic semiconductor, Nano Lett. **21**, 7050 (2021).

[55] N. Ding, K. Yananose, C. Rizza, F. R. Fan, S. Dong, and A. Stroppa, Magneto-optical Kerr effect in ferroelectric antiferromagnetic two-dimensional heterostructures, ACS Appl. Mater. Interfaces **15**, 22282 (2023).

[56] K. Yang, W. Hu, H. Wu, M.-H. Whangbo, P. G. Radaelli, and A. Stroppa, Magneto-optical Kerr switching properties of (CrI3)2 and (CrBr3/CrI3) bilayers, ACS Appl. Electron. Mater. **2**, 1373 (2020).

[57] N. Sivadas, S. Okamoto, and D. Xiao, Gate-controllable magneto-optic Kerr effect in layered collinear antiferromagnets, Phys. Rev. Lett. **117**, 267203 (2016).

[58] B. Pan, P. Zhou, P. Lyu, H. Xiao, X. Yang, and L. Sun, General stacking theory for altermagnetism in bilayer systems, Phys. Rev. Lett. **133**, 166701 (2024).

[59] S. Zeng and Y.-J. Zhao, Bilayer stacking A-type altermagnet: A general approach to generating two-dimensional altermagnetism, arXiv:2407.15097.